\pdfoutput=1

\documentclass[final,5p,times,twocolumn,authoryear]{elsarticle}

\usepackage{amssymb}
\usepackage{amsmath}
\usepackage{makecell}

\usepackage{hyperref}

\journal{Expert Systems With Applications}

\begin{document}

\begin{frontmatter}



\title{Urban Priority Pass: Fair Signalized Intersection Management \\ Accounting For Passenger Needs Through Prioritization }


\author{
\href{kriehl@ethz.ch}{Kevin Riehl}, \href{kouvelas@ethz.ch}{Anastasios Kouvelas}, \href{michail.makridis@ivt.baug.ethz.ch}{Michail A. Makridis}
} 

\affiliation{
        organization={Traffic Engineering Group, Institute for Transport Planning and Systems, ETH Zurich},
        addressline={Stefano-Franscini-Platz 5}, 
        city={Zurich},
        postcode={8093}, 
        state={Zurich},
        country={Switzerland}
    }

\begin{abstract}

Over the past few decades, efforts of road traffic management and practice have predominantly focused on maximizing system efficiency and mitigating congestion from a system perspective. 
This efficiency-driven approach implies the equal treatment of all vehicles, which often overlooks individual user experiences, broader social impacts, and the fact that users are heterogeneous in their urgency and experience different costs when being delayed.
Existing strategies to account for the differences in needs of users in traffic management cover dedicated transit lanes, prioritization of emergency vehicles, transit signal prioritization, and economic instruments.
Even though they are the major bottleneck for traffic in cities, no dedicated instrument that enables prioritization of individual drivers at intersections.
The Priority Pass is a reservation-based, economic controller that expedites entitled vehicles at signalized intersections, without causing arbitrary delays for not-entitled vehicles and without affecting transportation efficiency de trop.
The prioritization of vulnerable road users, emergency vehicles, commercial taxi and delivery drivers, or urgent individuals can enhance road safety, and achieve social, environmental, and economic goals.
A case study in Manhattan demonstrates the feasibility of individual prioritization (up to 40\% delay decrease), and quantifies the potential of the Priority Pass to gain social welfare benefits for the people. 
A market for prioritization could generate up to 1 million \$ in daily revenues for Manhattan, and equitably allocate delay reductions to those in need, rather than those with a high income.

\end{abstract}







\end{frontmatter}




\section{Introduction}
\label{sec:introduction}

Transportation infrastructure plays a crucial role in ensuring a meaningful, joyful, fulfilling, and dignified life in today’s society by providing access to the job market, educational and recreational activities, social participation, and marketplaces (\citealp{karner2020transportation}). 
Many people world wide are motorized vehicle (car) dependent, as road transportation often is the only accessible and affordable means of transportation, contrary to public transport which is often underdeveloped and not a competitive option (\citealp{sierra2024we,saeidizand2022revisiting}). 
Road infrastructure is shared by many users, and the utility of the infrastructure depends on the number of users; when being used over capacity, congestion arises, delays increase, and the utility for users diminishes. 
Ever growing population and demand for mobility, and the over-consumption of shared transportation infrastructure is a growing issue worldwide, leading to conflicts between the users resulting in congestion (\citealp{verhoef2010economics}).
As a result, previous efforts have gone into finding ways of distributing mobility resources in a fair and efficient way to mitigate these conflicts (\citealp{waller2025mobility,ni2024bicycle,livingston2023bike,riehl2024towards}).
In the urban context, it is especially the intersections that are the major bottleneck, causing significant delays during urban traffic peak hours (\citealp{hurdle1984signalized,wolshon1999analysis}).

Over the past few decades, transportation research and practice have predominantly focused on maximizing system efficiency and mitigating congestion from a macro-level perspective. 
This approach has prioritized overall traffic flow and network performance, often overlooking individual user experiences and broader societal impacts. 
Metrics such as throughput, queue length, average vehicle delay times, or total travel time are considered during system design and optimization. 
While this system-centric view has led to significant improvements in traffic management and average user's experience, it has sometimes neglected the human element of transportation, including factors such as user satisfaction, accessibility, and quality of life (\citealp{de2011handbook,goodwin1997solving}).
Moreover, a purely efficiency-driven approach not only generates beneficiaries among the users; misspecified designs of traffic engineering solutions harbor the threat of creating systematic inequalities between the users, unacceptably long waiting times, or excessive queue forming. 
Transportation efficiency is a multidimensional concept, and optimizing for one goal (e.g. reducing delays), can deteriorate conflicting goals (e.g. more pollution, more energy consumption). 

The efficiency-driven approach implies the equal treatment of all vehicles, which overlooks, that users are not equal; they are heterogeneous in their urgency and experience different costs when being delayed.
Experienced delay times can be considered as resource of negative utility for the users (\citealp{waller2025mobility}).
Depending on their need (urgency), users will experience a negative utility (cost) because of their experienced delays (\citealp{jacquillat2022predictive}).

The recognition of equity issues in transportation has catalysed a paradigm shift towards sustainable, equitable, and fair transport systems. 
This transformation acknowledges that fairness is a complex, intangible concept that requires striking a meaningful and justifiable balance in distributional questions.
A critical examination of the transportation literature reveals a tendency to equate equality with fairness, an oversimplification that fails to account for the nuanced nature of fairness and overlooks important differences among individuals and communities. 
In reality, the concept of fairness encompasses a spectrum of philosophical approaches beyond mere equality accounting for differences between users to better serve diverse communities and their unique needs
(\citealp{martens2016transport,randal2020fairness,riehl2024towards}). 

\begin{figure} [!htbp]
    \centering
    \includegraphics[width=0.98\linewidth]{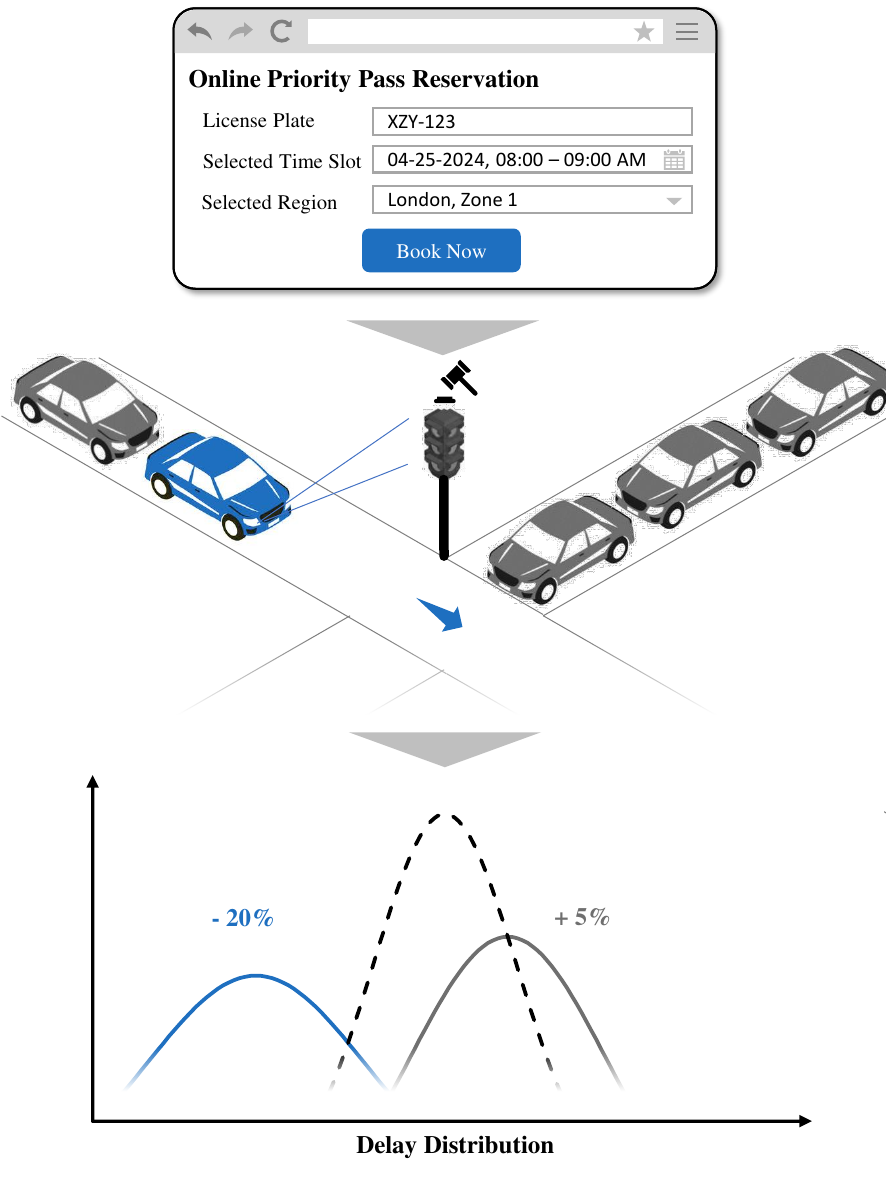}
    \caption{Urban Priority Pass Concept}
    \label{fig:priority_pass_concept}
\end{figure}

In order to account for the differences in needs between road users, prioritization is a commonly used approach in road traffic management.
In the urban context, the prioritization of public transport and emergency vehicles, as servers for the common good, is well-accepted and widely adopted.
While individual prioritization exists in highway contexts in the form of economic instruments in many countries via tolled lanes or roads, it can be rarely observed in urban contexts in simplistic, premature forms such as license plate rationing and congestion pricing, as side effects of traffic demand management.

Even though intersections are the major bottleneck in cities and the major source for delays, only few works explore the potential of prioritization at intersections in the urban context.
Contrary to other life contexts related to services where systems are designed to account for differences in needs between users (e.g., priority lanes when flying, express deliveries when online shopping, triage prioritization during medical treatment), mobility in urban traffic management fails to do so. 
Therefore, a form of prioritization to support those in need over those who are not so urgent could generate a significant value for the society.



This work sets out to contribute to the growing branch of literature on equitable transportation, individual signal prioritization, and reservation-based transportation systems, by proposing a novel, feasible, economic instrument for the urban context in order to address needs-oriented signalized intersection management in cities: the Urban Priority Pass, as shown in Fig.~\ref{fig:priority_pass_concept}.
Vehicles entitled with the Priority Pass shall receive privileged treatment at intersections and experience significantly shorter delays compared to not-entitled vehicles.
A certain amount of Priority Passes could be allocated to urgent vehicles for specific time slots, and considered in the control algorithms of signalized intersections.
Leveraging the non-linearities of near free-flow-traffic, the prioritization enables urgent drivers to influence their experienced delays while adequately compensating others, and overall leaves everyone better off due to the alignment of mobility resources with needs, social welfare gains, and potential additional revenues for the municipal administrations.

This work presents the concept of the Priority Pass, designs signal control algorithms based on auctions, analyses whether entitled vehicles can be expedited without causing arbitrary delays for the not-entitled vehicles, and if so, determines how many vehicles should be entitled optimally.
In addition to that, an assessment of the effects on transportation efficiency is conducted, and measures to quantify the generated welfare for the society by this form of prioritization are suggested.
A case study on a Manhattan-like map showcases the Urban Priority Pass concept and demonstrates its usefulness.
Afterwards, a discussion on fair and efficient ways of allocating the entitlement to drivers discourses how to ensure that users that are most in need are receiving privileged treatment. 
The results demonstrate, that prioritization is possible without affecting transportation efficiency de trop and without causing arbitrary, disproportionate delay increases for other drivers. 
Moreover, the results show that a market for prioritization expedites those in need rather than those with a high income, and highlights the potential welfare gains and revenues for municipalities (in social, environmental, and economic terms).

The rest of the paper is organized in following manner.
Section~\ref{sec:relatedworks} reviews related works on signalized intersection management, research on auction-controlled traffic lights, economic concepts (needs, market mechanisms, resources), discusses current approaches to need-orientation in road transportation, and highlights connections to reservation-based transportation systems and transit signal prioritization.
Section~\ref{sec:methods} outlines the functioning of the Urban Priority Pass, discusses how to measure the generated welfare, and describes the simulation and evaluation procedure into more detail.
The results of the case study are presented and discussed in Section ~\ref{sec:results}.
The work concludes with a summary, and an outlook to future research~\ref{sec:conclusion}.

\begin{figure*} [!htbp]
    \centering
    \includegraphics[width=0.98\linewidth]{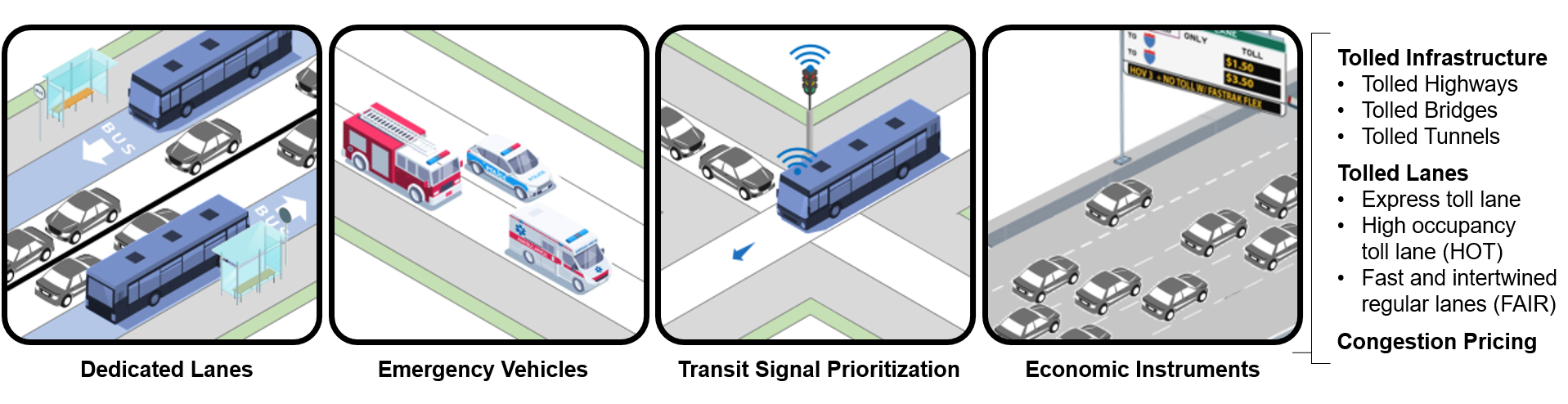}
    \caption{Vehicle Prioritization in Road Transportation}
    \label{fig:prioritization_example}
\end{figure*}

\section{Related Works}
\label{sec:relatedworks} 

This section motivates and highlights the Priority Pass concept by reviewing related literature from prioritization in road transportation, reservation-based transportation systems, and signalized intersection management.
Furthermore, a comparative overview of auction-controlled traffic lights demonstrates the potential to account for user preferences.
This section concludes with economic foundations such as needs, markets, auctions, and welfare, which is enables the design of benefit measures of the Priority Pass concept.

\subsection{Prioritization in road transportation systems}

Prioritization in road transportation systems is crucial for improving overall efficiency, for achieving strategic planning objectives, helps creating more sustainable, equitable, and liveable urban environments, and enables to account for differences in users. At the same time, prioritization can be challenging due to spatial or legislative constraints, and is reported to cause disruptions to the general traffic flow and to cause even more congestion overall under certain conditions (\citealp{turochy2001prioritizing}).
Generally, four groups of strategies to prioritize specific users in road transportation can be distinguished (see Fig.~\ref{fig:prioritization_example}):

(1) One can offer a dedicated share of the infrastructure to specific users, for example high occupancy vehicles (e.g., HOV lanes on highways), or public transport vehicles (e.g., bus rapid transit (BRT) lanes in urban contexts).

(2) One can prioritize emergency vehicles using road legislation. 
Ambulances, police cars, and fire trucks for instance, are prioritized in case of urgency to arrive their destination with a minimal delay.

(3) One can prioritize public transport vehicles at intersections (transit signal prioritization).

(4) One can apply economic instruments, namely road pricing, that enable users to trade-off time and monetary resources in market-like systems. 
This can include the tolling of dedicated parts of the infrastructure (e.g., tolled highways, bridges, tunnels), and pricing of dedicated lanes (e.g., express toll lanes, high occupancy toll (HOT) lanes, fast and intertwined regular (FAIR) lanes).

While the dedication of infrastructure to car-poolers or the public transport has many positive effects on overall traffic and transit level of service, space is often a scarce resource, especially in urban settings (\citealp{levinson2002bus, daganzo2008effects}). 

The prioritization of emergency vehicles enables a fast response for critical, public services, but at the same time can cause severe disruptions of traffic flows, especially at intersections in the urban context (\citealp{younes2018efficient}). As a consequence, a growing branch of work is dedicated to emergency vehicle preemption in traffic light control to anticipate disruptions and mitigate these (\citealp{qin2012control,djahel2013adaptive}).

Transit signal prioritization is a traffic management strategy that gives preferential treatment to public transport vehicles at signalized intersections. 
This can involve extending green lights, shortening red lights, or providing special signal phases for transit vehicles (\citealp{garrow1999development}). 
While the delays of public transport vehicles can be significantly reduced, there are also negative side effects such as a disruption of signal coordination, potential for increased overall delay, and limited flexibility under high congestion regimes (\citealp{lin2015transit}).
The novelty of the Priority Pass lies in that it enables a signal prioritization for almost any vehicle.

Transport demand management using economic instruments, such as congestion pricing, can be considered as prioritization as well. 
In this case, entering to or driving in a specific area is charged, which enables market-like management of infrastructure access and thus an allocation of travel times (prioritization) and costs according to needs (\citealp{bento2020avoiding}).
Often, the implementation of transport demand management measures is challenging due to public resistance, and the lack of support in political decision making processes, as economic instruments are considered to cause systematic equity issues (\citealp{riehl2024towards}).

\subsection{Reservation-based transportation systems}

In the recent decades, traffic congestion has emerged as a major challenge to road transportation systems both in freeway and urban contexts.
While investments into more road infrastructure and increasing road capacity can help to increase the supply on the short-term, demand is observed to grow on the long-term, resulting in even worse traffic situations (induced demand).
Therefore, various efforts went into aforementioned transport demand management strategies.
Reservation-based transportation systems are a concept of transport demand management, that allocate spatio-temporal mobility resources according to reservations that travellers need to conduct in advance (\citealp{edara2008model,lee1999induced}).

While reservations are common practice in public transport such as long-distance railways, buses, flights, or car sharing (\citealp{mitev1996more, copeland1988airline, gaggero2019pricing}), a growing branch of research explores the potential for reservations of single lanes (\citealp{liu2011token,liu2013design}), reservations for the usage of highways (\citealp{su2013proof,edara2008model}), reservations for parking spaces (\citealp{wang2011reservation}), reservations for specific routes (\citealp{makridis2021implementation}), and reservations for driving into cities using rationing (\citealp{nie2017license}) and tradeable credit schemes (\citealp{provoost2023design}).
The novelty of the Priority Pass lies in that it enables the reservation of prioritization.

\begin{figure*} [!htbp]
    \centering
    \includegraphics[width=0.98\linewidth]{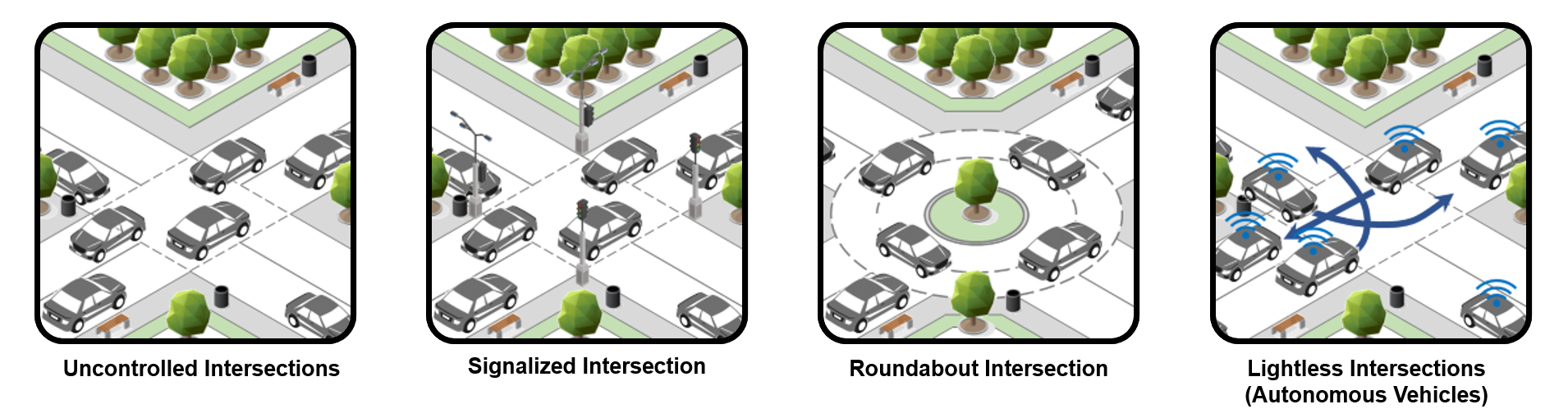}
    \caption{Road Intersection Management}
    \label{fig:intersection_example}
\end{figure*}

\subsection{Signalized intersection management}

Intersections are the major source for congestion in urban traffic networks, and mainly four designs can be found to solve conflicts at intersection as shown in Fig.~\ref{fig:intersection_example}.
There are uncontrolled intersections, that rely on cooperation and self-coordination of drivers, such as intersections of side streets. 
There are roundabout intersections, that are usually chosen for safety reasons over uncontrolled intersections.
Finally, there are controlled intersections, that can achieve higher capacities and are used at the intersection of critical, main traffic roads.
Traffic lights represent today’s predominant solution for intersection management (\citealp{wu2022intersection}).

The control algorithms of conventional traffic lights can be characterized by fixation of cycle, adaptivity, actuation, coordination, and connectivity (\citealp{qadri2020state}).
Non-conflicting passage routes at an intersection are grouped to sets of movement phases, which are given right of way by the traffic light for a certain green phase duration.
The goal of traffic light control is the planning of signal phase and timing (SPaT) plans, that achieve higher levels of efficiency, for example measured in number of vehicles per time unit passing (throughput).

The simplest forms of signal control implement a fixed-cycle of movement phases and related green phase durations that are repeatedly signalized. These fixed-cycle SPaTs can also be optimized for different periods of the day based on historical data.
Adaptive signal control improves these simple fixed-cycle controls, by adjusting green phase durations according to recent changes in traffic flows in real-time (\citealp{kouvelas2014maximum}).
In addition to that, actuated traffic light control not only adjusts the duration of green lights, but also chooses the next movement phase dynamically rather than following a specific order, in order to meet movement phases' demand for vehicle passage.
What's more, coordinated traffic lights take the traffic state at multiple intersections into consideration to achieve a better traffic performance on a network level.
Future signal control could even go a step further, and leverage vehicle-to-infrastructure communication to achieve higher capacities (\citealp{gao2020review}).
For fully-connected and fully-autonomous vehicles, lightless intersections could leverage vehicle-to-vehicle communication and achieve even higher capacities with advanced, automated self-coordination at intersections (\citealp{chavoshi2021pairing}).

\subsection{Auction-controlled traffic lights}

Traffic lights can be seen as controllers, that allocate spatio-temporal, right-of-way resources in the form of SPaTs to competing streams of vehicles, which motivates an economic perspective on signalized intersection management.
Therefore, auctions, amongst other market mechanisms for resource allocation, have been studied in a growing branch of literature (\citealp{iliopoulou2022survey}).
The proposed Priority Pass is designed as an auction-based controller for signalized intersection management, as this design enables to account for differences in users.

Even though, there are few, theoretical studies (\citealp{paperG,paperK,paperL,paperN}) that explore vehicles which can bid at intersections to reduce delays by paying depending on their needs, these require fully-connected vehicles and infrastructures, potentially autonomous vehicles, which is not feasible in the near future.
What's more, the decision making process (bidding) is assumed to happen at every intersection, which is computationally expensive.
Contrary to that, the proposed Priority Pass concept is a feasible solution that could be realized with current infrastructure at intersections, and facilitates the decision making process by a choice transfer from specific intersections to specific periods of times.

\begin{table*}[htbp!]
    \centering
    \begin{tabular}{|m{3.0cm}|l|l|l|l|l|l|m{3.0cm}|}
        \hline
        \scriptsize{\textbf{Ref. }} & \scriptsize{\textbf{Resource}} & \scriptsize{\textbf{Bidder}} & \makecell[bl]{\scriptsize{\textbf{System}} \\ \scriptsize{\textbf{Bidder}}} & \makecell[bl]{\scriptsize{\textbf{Auction}} \\ \scriptsize{\textbf{Trigger}}} & \makecell[bl]{ \scriptsize{\textbf{Payment}} \\ \scriptsize{\textbf{System}}} & \scriptsize{\textbf{Simulator}} & \scriptsize{\textbf{Map}} \\
        \hline
        
        \scriptsize{\cite{paperA}} & \scriptsize{Next Phase}     & \scriptsize{Phase}   & \scriptsize{Yes} & \scriptsize{Phase}     & \makecell[tl]{\scriptsize{Money} \\ \scriptsize{(2nd highest bid)}} & \scriptsize{AORTA}    & \scriptsize{Austin, Baton Rouge, San Francisco, Seattle} \\
        & & & & & & & \\
        \scriptsize{\cite{paperB}} & \scriptsize{Phase Duration} & \scriptsize{Phase}   & \scriptsize{Yes} & \scriptsize{Intervals} & \scriptsize{(None)} & \scriptsize{SUMO}     & \scriptsize{Manhattan-like (4x4)} \\
        & & & & & & & \\
        \scriptsize{\cite{paperC}} & \scriptsize{Next Phase}     & \scriptsize{Phase}   & \scriptsize{No}  & \scriptsize{Intervals} & \scriptsize{(None)} & \scriptsize{SUMO}     & \scriptsize{Mountain-View (California)} \\
        & & & & & & & \\
        \scriptsize{\cite{paperD}} & \scriptsize{Next Phase}     & \scriptsize{Phase}   & \scriptsize{No}  & \scriptsize{Intervals} & \makecell[tl]{\scriptsize{Token} \\ \scriptsize{(highest bid)}} & \scriptsize{(?)}      & \scriptsize{Sythetic (6 intersections)} \\
        & & & & & & & \\
        \scriptsize{\cite{paperE}} & \scriptsize{Phase Duration} & \scriptsize{Phase}   & \scriptsize{Yes} & \scriptsize{Cycle}      & \scriptsize{(None)} & \scriptsize{AIMSUN}   & \scriptsize{Porto (single intersection)} \\
        & & & & & & & \\
        \scriptsize{\cite{paperF}} & \scriptsize{Phase Duration} & \scriptsize{Phase}   & \scriptsize{No}  & \scriptsize{Phase}      & \scriptsize{(None)} & \scriptsize{SUMO}     & \scriptsize{Chicago} \\
        & & & & & & & \\
        \scriptsize{\cite{paperG}} & \scriptsize{Phase Duration} & \scriptsize{Vehicle} & \scriptsize{No}  & \scriptsize{Phase}      & \makecell[tl]{\scriptsize{Money} \\ \scriptsize{(highest bid)}} & \scriptsize{(Python)} & \scriptsize{Synthetic (1 intersection)} \\
        & & & & & & & \\
        \scriptsize{\cite{paperH}} & \scriptsize{Phase Duration} & \scriptsize{Phase}   & \scriptsize{Yes} & \scriptsize{Intervals}  & \scriptsize{(None)} & \scriptsize{SUMO}     & \scriptsize{Manhattan-like (4x4)} \\
        & & & & & & & \\
        \scriptsize{\cite{paperI}} & \scriptsize{Next Phase}     & \scriptsize{Phase}   & \scriptsize{No}  & \scriptsize{Phase}      & \scriptsize{(None)} & \scriptsize{VISSIM}   & \scriptsize{Synthetic (15 intersections)} \\
        & & & & & & & \\
        \scriptsize{\cite{paperJ}} & \scriptsize{Phase Duration} & \scriptsize{Phase}   & \scriptsize{No}  & \scriptsize{Phase}      & \makecell[tl]{\scriptsize{Money} \\ \scriptsize{(highest bid)}} & \scriptsize{ASEP}     & \scriptsize{Synthetic (1 intersection)} \\
        & & & & & & & \\
        \scriptsize{\cite{paperK}} & \scriptsize{Phase Duration} & \scriptsize{Vehicle} & \scriptsize{No}  & \scriptsize{Phase}      & \makecell[tl]{\scriptsize{Token} \\ \scriptsize{(difference 2} \\ \scriptsize{highest bids})} & \scriptsize{(Java)}   & \scriptsize{Synthetic (1 intersection)} \\
        & & & & & & & \\
        \scriptsize{\cite{paperL}} & \scriptsize{Phase Duration} & \scriptsize{Vehicle} & \scriptsize{No}  & \scriptsize{Phase}      & \makecell[tl]{\scriptsize{Token} \\ \scriptsize{(highest bid)}} & \scriptsize{MATSIM}   & \scriptsize{Modena (MASA area)} \\
        & & & & & & & \\
        \scriptsize{\cite{paperM}} & \scriptsize{Phase Duration} & \scriptsize{Phase}   & \scriptsize{No}  & \scriptsize{Phase}      & \scriptsize{(None)} & \scriptsize{(?)}      & \scriptsize{Synthetic (1 intersection)} \\
        & & & & & & & \\
        \scriptsize{\cite{paperN}} & \scriptsize{Phase Duration} & \scriptsize{Vehicle} & \scriptsize{No}  & \scriptsize{Phase}      & \makecell[tl]{\scriptsize{Token} \\ \scriptsize{(highest bid)}} & \scriptsize{MATSIM}   & \scriptsize{Modena (MASA area)} \\
        & & & & & & & \\
        \hline
    \end{tabular}
    \caption{Literature Review: Auction-controlled, Signalized Intersection Management}
    \label{tab:aucSigInt}
\end{table*}

We provide an exhaustive review of all works that apply auctions in the context of signalized intersection management in Table~\ref{tab:aucSigInt}.
Auctions can be used as adaptive or actuated traffic light controllers; one study even investigates coordinated intersection management with auction-based signal control (\citealp{paperI}).
The works applying auction-based control mainly differ in which resources are auctioned by which auction participants (bidders), how much they bid, whether they pay for, who receives the payment, what kind of payment system is used, whether there is a system bidder participating as well, and how auctions are triggered over the course of time. 

\begin{figure*} [!htbp]
    \centering
    \includegraphics[width=0.98\linewidth]{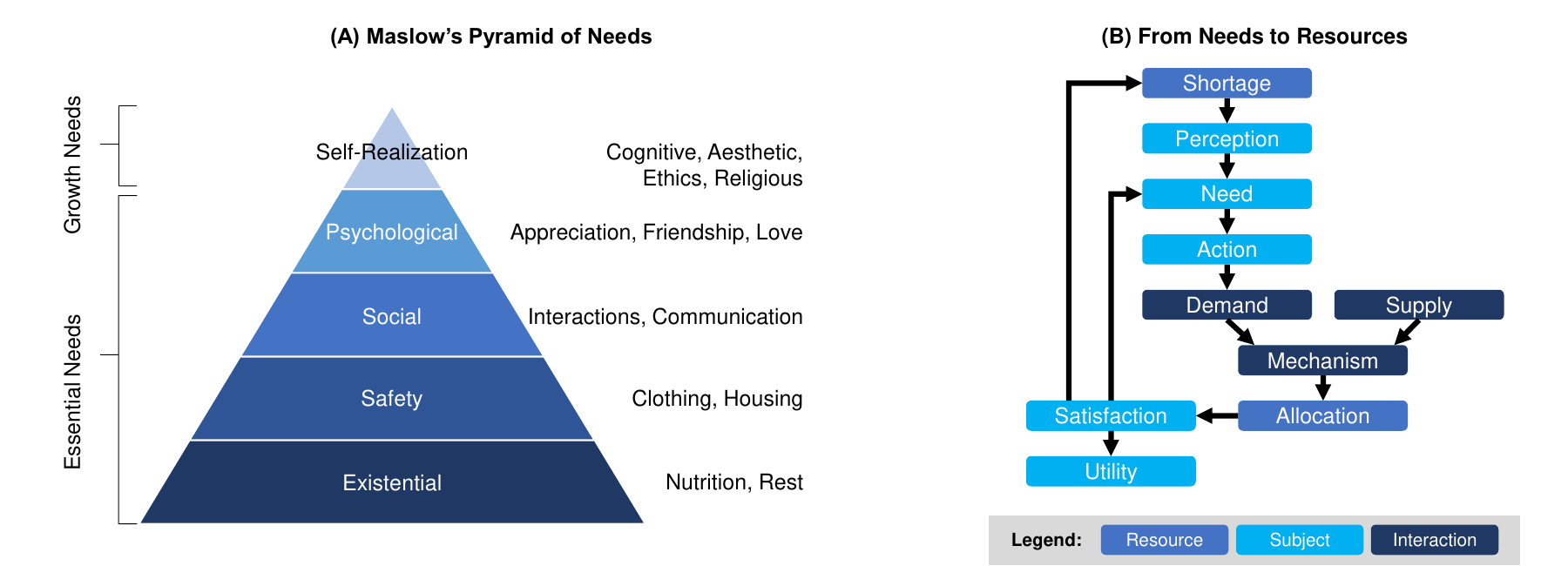}
    \caption{Needs Theory in Psychology and Economics}
    \label{fig:need_theory}
\end{figure*}

There are mainly two types of resources auctioned in the studies, either the next movement phase or the change of phase durations.
Amongst the bidders, most studies focus on movement phase bidders, that aggregate bids and represent the interest of vehicles in all lanes related to a movement phase. 
Besides, there are studies that discourse single vehicle bidders (either all, or just the front vehicle at queues).
Besides, several studies suggest the introduction of a system bidder, that intervenes with bids in addition to the auction participants, to guarantee a maximum red, and a minimum green time for each movement phase.
The auction can either be triggered on a regular interval basis, after a whole fixed-cycle passed, or after a movement phase's green duration passed.
Most works apply a sealed-bid, first-price, single-item alike auction, where the highest bid, the second highest bid, or the difference between highest and second highest bid need to be paid. 
Several works suggest a monetary payment system, while others propose virtual currency, token-based systems such as "Virtual Coins" (\citealp{paperL,paperM}) and "Transportation Points" (\citealp{paperK}). 
A large share of works also suggests no payments at all (e.g. in the case of movement phase bidders that count the vehicles in their lanes rather than aggregating their bids).

The bidding strategy (policy) of auction participants is analysed in various different ways.
A major determinant of bidding strategy is whether the participants act cooperative or non-cooperative; in case of single vehicle bidders one must assume the first, and in case of movement phase and system bidder one can also assume the latter.
Several works propose a dedicated algorithm (\citealp{paperA,paperB,paperC,paperE,paperH}), define a budget per intersection (\citealp{paperL,paperN}), or compute an optimal solution using game theoretic models and optimization (\citealp{paperD,paperF,paperG,paperI,paperJ,paperK,paperM}).

\subsection{Needs, Urgency, and Welfare}


Resources are vital for a good life and well-being of humans.
Resources can be defined as (material \& immaterial) goods that have a utility to people, as resources satisfy their needs.
A need can be seen as a subjective state preference.
For instance, the preference of a nutritioned over a malnutritioned state is an essential need for humans.
Needs are the major driver for human motivation, which is why it has been extensively studied in philosophy, psychology, and economics (\citealp{wheeler1980basic,kamenetzky1981economics}).

There are many different types and models to classify needs, where Maslow's need pyramid is one of the most renowned models and depicted in Fig.\ref{fig:need_theory}(A).
\citep{maslow1943theory} categorised needs in a pyramid-scheme model of five levels.
The needs of the first four layers can be considered as deficient or essential needs.
These types of need have in common that they can be saturated, and do not require infinite amounts of resources.
The absence of satisfaction of these needs is causing deficits in well-being and happiness.
An important observation to be made is, while the lower layers relate to material resources, the upper layers relate more to immaterial resources, and can not necessarily be saturated.

The relationship between needs and resources is summarized in Fig.\ref{fig:need_theory}(B).
Shortage of a resource is a state.
Depending on subjective preferences, this state can be perceived as undesirable.
As a result of the subjective perception process, a need is generated.
This need leads to an individual action.
All actions aggregated together can be considered as the demand of a population.
This demand meets a supply (where we do not specify the origin of the supply any further here).
A mechanism transforms demand and supply into an allocation of resources to the population.
The allocated resources satisfy the shortage and needs of individuals, which represents utility for the individuals.

Economics is the study of production, consumption, and allocation of resources.
Markets are resource allocation mechanisms in which two groups, the demanders and suppliers, meet.
In barter markets, participants exchange resources directly, while in monetary markets a commonly accepted mean of payment (money) is the object of exchange.
Needs lead to a willingness to pay a certain price (action).
Efficient resource allocation describes the distribution of resources that maximizes the population's utility.
The fundamental theorems of welfare economics describe the necessary conditions for markets to be efficient; this includes inter alia perfect information, competition and symmetry of power to act (\citealp{blaug2007fundamental}).

Auctions are a type of market, in which participants place bids and depending on the type of auction, an allocation rule determines which participants conduct a transaction. 
In English auctions, demanders place increasing bids until suppliers agree to sell for a certain price. 
In Dutch auctions, suppliers place decreasing bids until demanders agree to buy for a certain price. 
Contrary to open auctions, participants do not have transparency over the bids of their peers in sealed-bid auctions.
In single-shot auctions, there is only one round of bid placement, while in multi-shot auctions multiple rounds are possible.
In first-price auctions, the bidder with the highest bid wins, while in second-price auctions the bidder with the second highest bid wins.
After determining an auction winner, various payment rules are possible, such as that the winner pays its bid, the difference between its and the second-rank winner, the bid of the second-rank winner, etc.
It was shown, that a sealed-bid, second-price auction, the so called Vickrey auction, incentivises bidders to bid their true value (incentive compatibility), which is one of the requirements for efficient resource allocation (\citealp{krishna2009auction}).

In the context of signalized intersection management, spatio-temporal, network-related, mobility resources(~\citealp{waller2025mobility}) are allocated to a population of users by a mechanism (the traffic light controller).
Delays can be considered the resource of interest, that satisfies the need of urgency.
Delays are temporal resources with a negative utility to the users, that occur when users cannot travel in the fastest possible way due to conflicts with others.
Urgency is a temporal need, where shorter times are preferred over longer (travel) times.
Experienced (allocated) delays cause a negative utility (cost) to the transportation network user depending on its urgency. 
At a conventional signalized intersection, it is not possible for the user to take any action to express its urgency to be prioritized in any form; users can do nothing but wait.
The Priority Pass, serves as the missing link of action between need and resource allocation in this context.

\section{Methods}
\label{sec:methods} 

In this section we present the concept and functioning of the Priority Pass traffic light controller into detail.
Next we discuss the benefit of the Priority Pass by developing a social welfare measure in the context of signalized intersection management.
Then we elaborate on the allocation of Priority Pass entitlement using a reservation system.
Finally, we present case study \& simulation environment that are used to evaluate the Priority Pass controller.

\subsection{Priority Pass Traffic Light Controller}

The proposed Urban Priority Pass employs auction-controlled, signalized intersection management.
The Priority Pass concept employs a single-shot, sealed-bid, first-price auction scheme, where movement phase bidders compete for the resource of the next selected phase in the absence of a system bidder and payment system. 
Each movement phase $p$ determines its bid $b_p$ based on the number of vehicles $n_p$, and the number of entitled vehicles~\footnote{The entitled vehicles could be recognized with vehicle identifying infrastructure, such as license plate recognition, front window stickers, or near field communication, which already is common practice on highways and larger cities.} $e_p$:
\begin{equation}
b_p = (1 - \tau) \times n_p + \tau \times e_p  \:\: , \: \tau \in [ 0; 1]
\end{equation}

The Priority Pass has mainly two design parameters: the control threshold $\tau$ and the entitlement share $\gamma$.
The control threshold $\tau \in [ 0; 1]$ represents how strongly the Priority Pass is taken into consideration by the control algorithm.
The larger $\tau$, the more the controller privileges \& expedites the entitled vehicles over the non-entitled vehicles.
The entitlement share $\gamma \in [0; 1]$ represents how many vehicles from the population are entitled with the Priority Pass.

The traffic light signalling works as follows.
Once a movement phase $p^{*}$ becomes green, it stays green for at least $t_{min}$.
Afterwards, auctions are triggered at regular intervals $t_{auc}$.
In case any movement phase $p^{°}$ exceeds the maximum red period duration $t_{max}$, a signal transition towards this phase is initiated to guarantee reasonable upper-bound waiting times for low-traffic phases.
In case the current green phase looses the auction, a signal transition towards the auction winner is initiated.
The signal transition from one phase to the next happens via an intermediate phase with yellow signals, that takes a duration of $t_{trans}$ to account for driving security.

\begin{figure*} [!htbp!]
    \centering
    \includegraphics[width=0.98\linewidth]{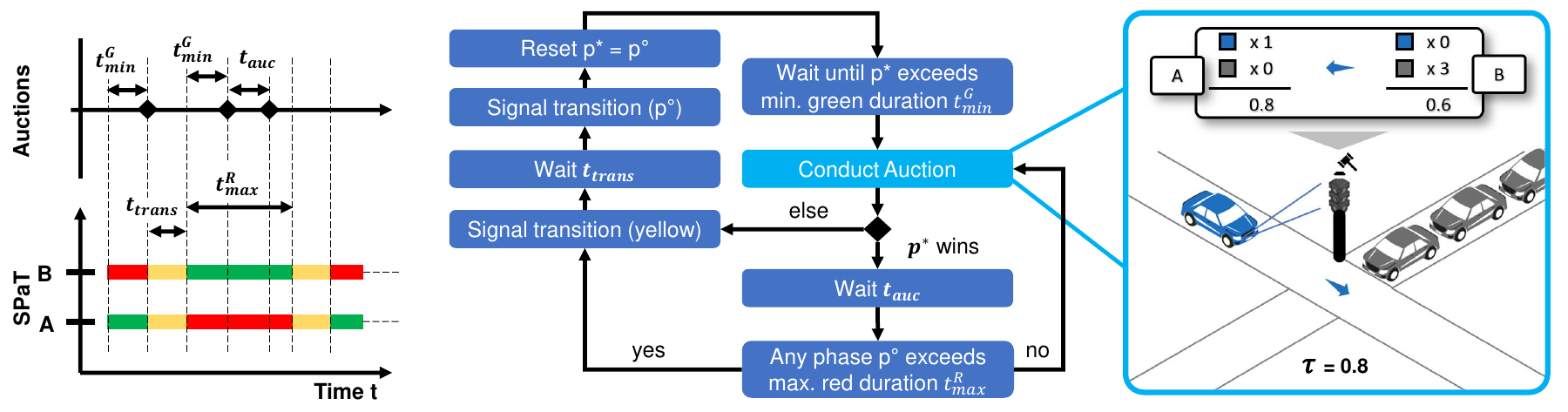}
    \caption{Priority Pass Traffic Light Auction-Controller}
    \label{fig:trafficlightcontroller_expl}
\end{figure*}

Fig.~\ref{fig:trafficlightcontroller_expl} summarizes the functioning of the Priority Pass signal control.
The left plot shows the SPaT and executed auctions across the time.
The plot in the middle describes the temporal logic of auction triggering.
At the beginning, phase A remains green for $t^{G}_{min}$. 
After that, an auction is triggered, resulting in a signal transition to phase B for $t_{trans}$.
After another $t^{G}_{min}$, two auctions are triggered, where phase B always wins, as it has higher number of vehicles.
After $t^{R}_{max}$ red time for phase A, a signal transition to phase A is initiated.
The right figure exemplifies an auction, where movement phase B has much more traffic than phase A, but A has one entitled vehicle (blue) and thus wins (for $\tau=0.8$).

The auction-based Priority Pass controller can be considered as a Max-Pressure controller (\citealp{varaiya2013max}) for $\tau = 0$ respectively $\gamma = 0$.
Depending on the available road sensor infrastructure, $n_p$ and $e_p$ could also represent queue length or link occupancies.
Within this study, following values are considered: 120 seconds for $t^{r}_{max}$, and 3 seconds for $t_{trans}$. 
The value for $t^{G}_{min}$ is determined based on optimization.

\subsection{Measuring the Benefit of the Priority Pass}

Vehicles that pass the road network experience an average delay per travelled distance $\delta$ [s/km].
This delay causes a cost $c$ [€/km] to the drivers.
This cost depends on their urgency $u$ [€/h]; for instance, an urgent ride to catch the next flight at the airport has higher delay costs than cruising for leisure purposes or to find a parking spot.
It is common to represent a vehicle's urgency by its value of time (VOT) (\citealp{moses1963value}).
The delay cost of a vehicle can be determined in monetary values as the product of delay and VOT $c = \delta \times u$.

Signalized intersection management that does not account for the differences in urgencies of the vehicles, causes equal delays for all vehicles $\delta_{avg}$ on average.
Contrary to that, the Priority Pass has the potential to allocate delays more efficiently according to the needs, and thus reduces total costs for all vehicles by prioritizing and expediting highly urgent, entitled vehicles.
The prioritization will create two groups of vehicles, and cause average delays for entitled $\delta_{PP} < \delta_{avg}$ and not-entitled $\delta_{NPP} > \delta_{avg}$ vehicles. 
Average urgencies $u_{PP}$ and $u_{NPP}$ are assumed for the two groups of vehicles, where $u_{PP} > u_{NPP}$.

The reduction in costs can be considered as the measure to quantify the benefit of the Priority Pass for the average user:
\begin{equation}
c_{r} = \gamma (\delta_{avg} - \delta_{PP}) u_{PP} + (1 - \gamma) ( \delta_{avg} - \delta_{NPP} ) u_{NPP} 
\end{equation}

The benefit of the Priority Pass for the system (total population) $C$ [€/h] can be quantified as the average vehicle's benefit $c_{r}$ multiplied with the total flow of vehicles $F$ [veh/h] and the average trip length $l_{avg}$ [km]:
\begin{equation}
C_{r} = c_{r} \times F \times l_{avg}
\end{equation}

When choosing the Priority Pass hyper parameters $\gamma$ and $\tau$, one must consider two trade-offs: (i) the delay reductions for prioritized vehicles shall not come at the cost of arbitrary delays for not-entitled vehicles, and (ii) the system's overall transportation efficiency shall not be affected by the introduction of the Priority Pass.
Complying with these two trade-offs, the benefits $c_{r}$ respectively $C_{r}$ serve as suitable goal functions when optimizing for finding suitable $\gamma$ and $\tau$.


\subsection{Reservation System For Allocation Of Priority Pass}

As mentioned in the introduction and review section, the novelty of the Priority Pass lies in that it conceptualizes a reservation system for prioritization, where the prioritization happens fully-automatic through vehicle identifying infrastructure and auctions with movement phase bidders at every intersection.
This reservation dedicates the right for prioritization during a specific time slot and within a specific (city) region.
Moreover, the aforementioned benefit measures might be useful to determine how many vehicles ($\gamma$) should be entitled, and how much these entitled vehicles should be prioritized ($\tau$).
However, the question of who should be prioritized, remains.

Possibly, Priority Pass reservations could be granted by a central authority to specific vehicles in order to meet environmental, social, or economic goals.
First, the authority could entitle public transportation vehicles for transit signal prioritization to incentivize use of public transportation, or blue light vehicles for emergency vehicle signal preemption. 
Second, the authority could grant the Priority Pass to vulnerable road users, such as cyclists and motorcycles, to incentivize active modes of transportation, and to enhance the safety of these road users.
Third, the authority could entitle drivers from low-income households, to advocate social equity, as these users usually live farther away, commute longer distances, and experience greater delays when commuting to work.
Fourth, the authority could grant the Priority Pass to delivery vehicles or taxis, as this would foster the municipal economy.
Independent of the underlying goal (social, environmental, or economic), such a centralized allocation conducted by an authority requires public debate and qualitative discussions, which is not straight forward in practice.

\begin{figure*} [!htbp]
    \centering
    \includegraphics[width=0.98\linewidth]{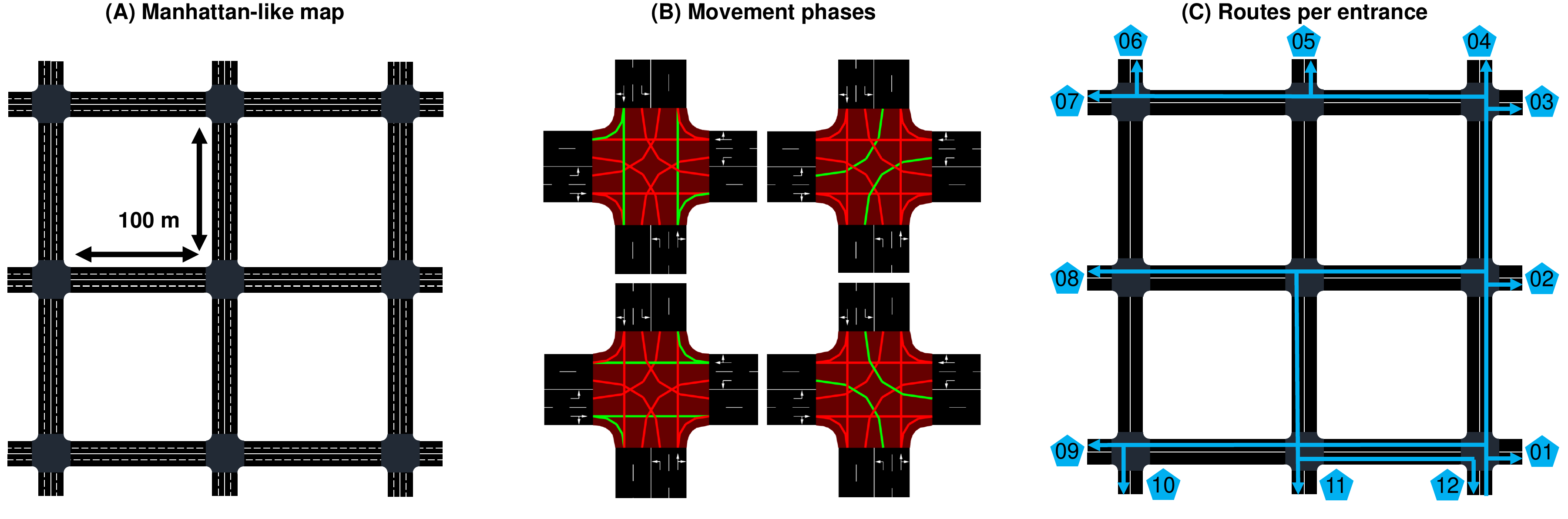}
    \caption{Manhattan-Like Case Study Map For Traffic Simulations}
    \label{fig:map}
\end{figure*}

Decentralized, market-based allocation could be a feasible alternative to redistributive, governmental activities.
As discussed before, the Priority Pass has the potential to reduce the costs of drivers by accounting for their urgency.
An important assumption made here, is that the urgency of entitled vehicles shall be higher compared to those that are not.
Market mechanisms can be an efficient tool to decentrally, self-coordinate the efficient (needs oriented) allocation of resources, as consumers are enabled to express their need (urgency) by a willingness to pay.
Therefore, we also foresee the possibility of users to buy the Priority Pass entitlement on a market.
Following economic theory and assuming efficient markets and rational users, those in highest need will be those willing to pay most, and thus receiving the Priority Pass entitlement.
For instance, taxi or delivery drivers would have a commercial interest to experience lower delays and thus would have a willingness to pay for this service.
In order to compensate those users that do not receive the Priority Pass adequately, we suggest that the payments of the beneficiaries could be distributed equally among the not-entitled users.
As a result, all users should be better off compared to the initial situation where no prioritization happens, as the Priority Pass enables drivers to actively influence their experienced delays and enables the traffic light controllers to account for differences of users.
The generated payments could not only be redistributed amongst the not-entitled users, but also (partially) collected as revenues for the municipal road authorities to foster public transportation and maintenance of roads.

\subsection{Case Study \& Simulation Design}

In order to demonstrate the usefulness of the Priority Pass, two types of simulations are conducted:
(i) traffic simulations to assess the effects on delays and traffic efficiency, and (ii) market simulations for the analysis of consumer behaviour to buy the Priority Pass.

\subsubsection{Traffic Simulations}
The case study employs a Manhattan-like map (Fig.~\ref{fig:map}A) that consists of nine signalized intersections each connected by bi-directional, four-laned (two lanes per direction), 100m-long roads.
Urban roads with speed limitations of 13.89 m/s (50 km/h) are assumed.
Each intersection is signalised with four distinct movement phases (Fig.~\ref{fig:map}B). 
The map is designed with a symmetric geometry; 
Fig.~\ref{fig:map}C shows the possible routes (destinations) of vehicles entering from a specific origin at the example of the bottom right entrance.

For computational simulation, the time-discrete \& space-continuous, microscopic traffic simulation environment SUMO (\citealp{SUMO2018}) with its standard settings for vehicles, and one second iteration steps, is used.
For simplicity reasons, an equal flow at each of the twelve entrances, that is varied from 50 to 550 vehicles per hour, and generated randomly with a exponential probability distribution ($\lambda=5$), is assumed.
Every time a vehicle is spawned, it is randomly assigned to a route from its entrance to one of the twelve exits with an even probability distribution. 
Each simulation consists of ten minutes simulation for transient response, before the recording of 60 minutes of simulation starts. 
Each simulation is repeated for ten times with different random seeds, and average of the metrics is calculated, in order to mitigate the random effects of the simulation and to account for the stochastic nature of urban traffic.

The Priority Pass controller is compared with two benchmark control algorithms in static contexts: the fixed-cycle control and the Max-Pressure controller.
This comparison allows to assess the impact of the Priority Pass when compared with both a non-actuated, and an actuated signal controller. 
Due to the static design of the simulation (a fixed flow per simulation), adaptive elaborations are put aside.
The fixed-cycle control gives the right of way to movement phases in a fixed, row-wise order as depicted in Fig.~\ref{fig:map}(B). 
Due to the symmetric design of a single intersection, equal green phase durations for movement phases of the same kind (straight\&left-turns, right-turns) are assumed.
Therefore, each intersection has two green-phase durations $t_f^1$ and $t_f^2$ that need to be optimized.
Due to the symmetric design of the whole network, equal green phase durations for each intersection are assumed. 
A time shift of $(t_f^1 + t_f^2)/2$ between the fixed cycles is applied in a chessboard-like pattern to account for a better flow between intersections.
Therefore, the fixed-cycle control is determined by the design parameters $t_f^1$ and $t_f^2$, which are optimized in a search range from 1 to 40 seconds.
The Max-Pressure controller is designed similar to the Priority Pass controller, where only the number of vehicles $n_p$ is taken into account when determining bids (implying $\gamma=0$ and $\tau=0$).
The Max-Pressure controller is determined by the design parameters $t_{min}$ and $t_{auc}$, which are optimized in a search range from 1 to 40 seconds.

Following measures are used to assess transportation efficiency: vehicle throughput, vehicle completion rate, average queue length, vehicle delay per kilometer, and the total travel time.
Both benchmark controllers are optimized for throughput, queue length, delay and total travel time.
One can observe that controllers that were optimized for total travel time, perform well across all efficiency metrics.
This is, why the final choice of the optimization goal function is set to the total travel time.

The Priority Pass controller is determined by the design parameters $\tau$ and $\gamma$, which are optimized in a search grid from 0.0 and 1.0 each.
All other, auction related parameters ($t_{min}$, $t_{auc}$) are undertaken from the optimized Max-Pressure benchmark controller.

\begin{figure} [!htbp]
    \centering
    \includegraphics[width=0.98\linewidth]{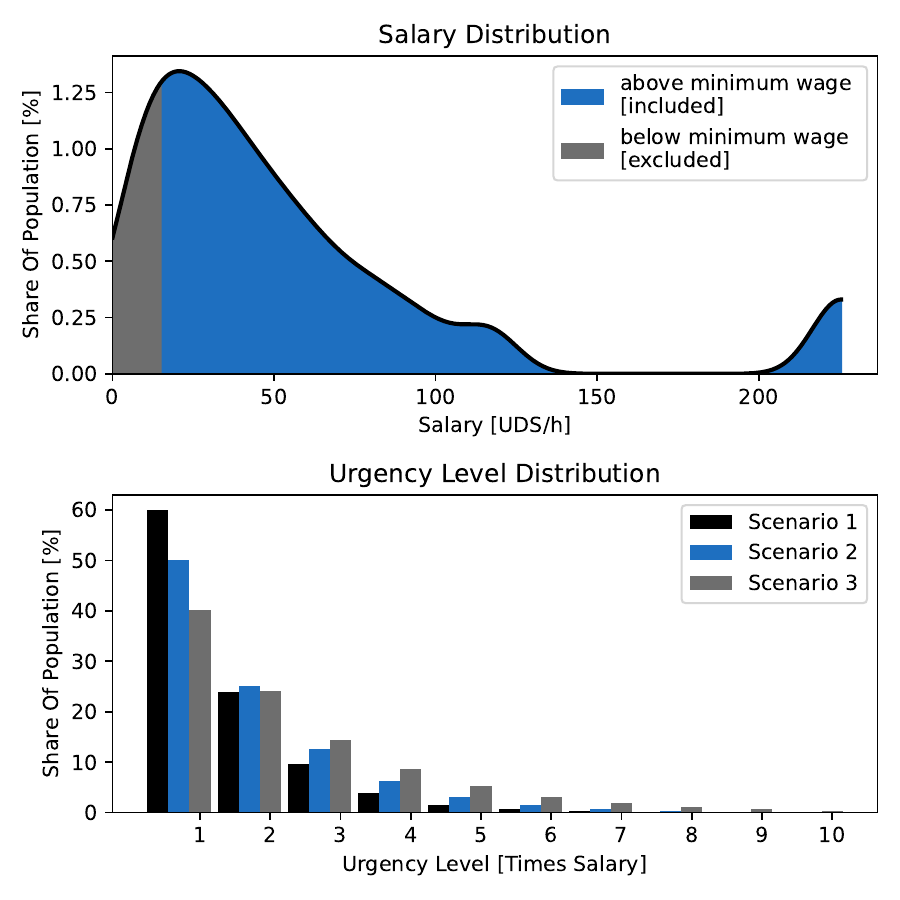}
    \caption{Market Model For Driver Population in New York City}
    \label{fig:market_model}
\end{figure}

\subsubsection{Market Simulations}

In order to simulate the consumer behaviour on a market for prioritization, we follow the market modelling of ~\citealp{riehl2024karma} for New York City.

Assume there are $n$ consumers, each with an hourly salary $w_i$, and a time-slot dependent urgency level $l_i$.
Assume a stochastic urgency processes that follow a geometric distribution ($P(X)=(1-p)^{x-1}p$), where larger shares of the population are not urgent, and smaller shares possess an increasing urgency level.
While the urgency level $l_i$ cannot be observed in markets, the willingness to pay $VOT_i$ (the value time has for the individuals) can.
The urgency level $l_i$ in \citealp{riehl2024karma} can be interpreted as the willingness to pay $x$-times the hourly salary to save one hour of time.
Therefore, a proportionality between $VOT_i$, $l_i$, and $w_i$ is assumed:
\begin{equation}
    VOT_i = l_i \times w_i
\end{equation}

We consider the real-world salary distribution of New York City according to 2022 US Census data,
and three different scenarios with varying urgency distribution ($p\in\{40\%, 50\%, 60\%\}$), as shown in Fig.~\ref{fig:market_model}.
Individuals with a hourly salary below minimum wage (15 \$/h) are excluded as these are assumed not to drive to work but rather use other modes of travel, such as public transportation.

\section{Results}
\label{sec:results} 

\begin{figure*} [!htpb]
   \centering
   \includegraphics[width=0.98\linewidth]{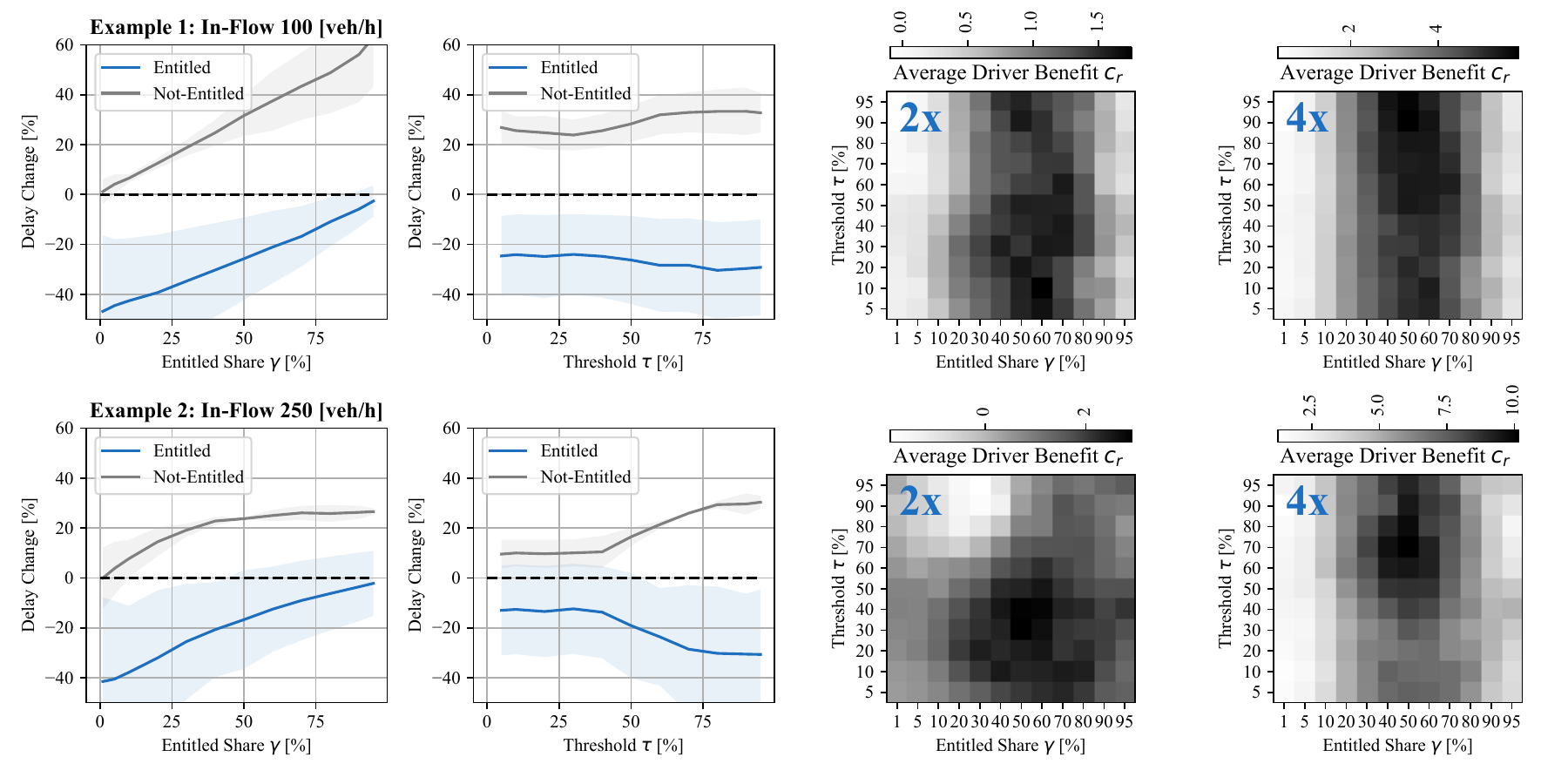}
   \caption{Mechanics of the Priority Pass}
   \label{fig:fig2_prioritypassmechanics}
\end{figure*}

In this section we discuss the results of this study, demonstrate the feasibility of the Priority Pass and its benefits.
First, we analyse the mechanics and delay distribution changes of the Priority Pass for different parameters $\gamma$ and $\tau$ at different traffic flow scenarios.
Second, we reveal, that for certain parameter combinations a significant benefit can be achieved through prioritization, and how the Priority Pass is optimized.
Third, we benchmark the Priority Pass against the fixed-cycle and the Max-Pressure controller, in order to assess whether the generated benefits come at the cost of significant, arbitrary transportation efficiency losses. Moreover, we analyse the effects of the Priority Pass controller on the green phase duration distribution of traffic lights.
Fourth, we conduct a market simulation study to better understand the value the Priority Pass could generate for a real-world scenario when being allocated with a decentralized, market-based mechanism.

\subsection{Applicability \& Mechanics of the Priority Pass}

Fig.~\ref{fig:fig2_prioritypassmechanics} illustrates the mechanics of the Priority Pass parameters $\gamma$ and $\tau$ for two different, static examples.
The examples differ in the number of vehicles per hour entering from each entrance of the network (100 veh/h vs. 250 veh/h).
The diagrams on the left outline the relative delay changes for prioritized (entitled with Priority Pass, blue) and not-prioritized vehicles (not entitled, gray).
Prioritized vehicles possess a negative delay change, meaning delays are reduced, while the not-prioritized vehicles possess a positive delay change, meaning delays are increased.
The more vehicles of the population get prioritized (larger $\gamma$), the smaller the benefits for prioritized and the greater the disadvantages for the not-prioritized vehicles.
The stronger entitled vehicles get prioritized (larger $\tau$), the greater the benefits for prioritized and the smaller the disadvantages for the not-prioritized vehicles.
When comparing the two traffic scenarios, one can observe that delay changes react more sensitive to $\gamma$ for smaller traffic flows, and that delay changes react more sensitive to $\tau$ for greater traffic flows.
The heatmaps on the right exhibit a potential average driver benefit for different Priority Pass parameters $\gamma$ and $\tau$ for the two traffic scenarios.
To calculate these average driver benefits, we make the simplifying assumption that the VOT of not-prioritized vehicles equals 1, and the VOT of the prioritized vehicles equals two times respectively four times the VOT of the not-prioritized vehicles.
The heatmaps outline that positive average driver benefits are possible through prioritization, and that the average driver benefit serves as a tool to optimize the Priority Pass (select suitable parameters $\gamma$ and $\tau$) so that the benefits of the prioritized vehicles outweighs the disadvantages of the not-prioritized vehicles, and not arbitrary cost arise.

To summarize, the results demonstrate, that an expedition is possible, significant delay reductions for prioritized vehicles are possible, and that the arbitrary delay increases for the not-prioritized vehicles are negligible (based on the assumptions of VOTs).
Moreover, the results yield insights into the mechanics of the Priority pass.

\subsection{Optimization of the Priority Pass Parameters}

\begin{figure*} [!htpb]
   \centering
   \includegraphics[width=0.98\linewidth]{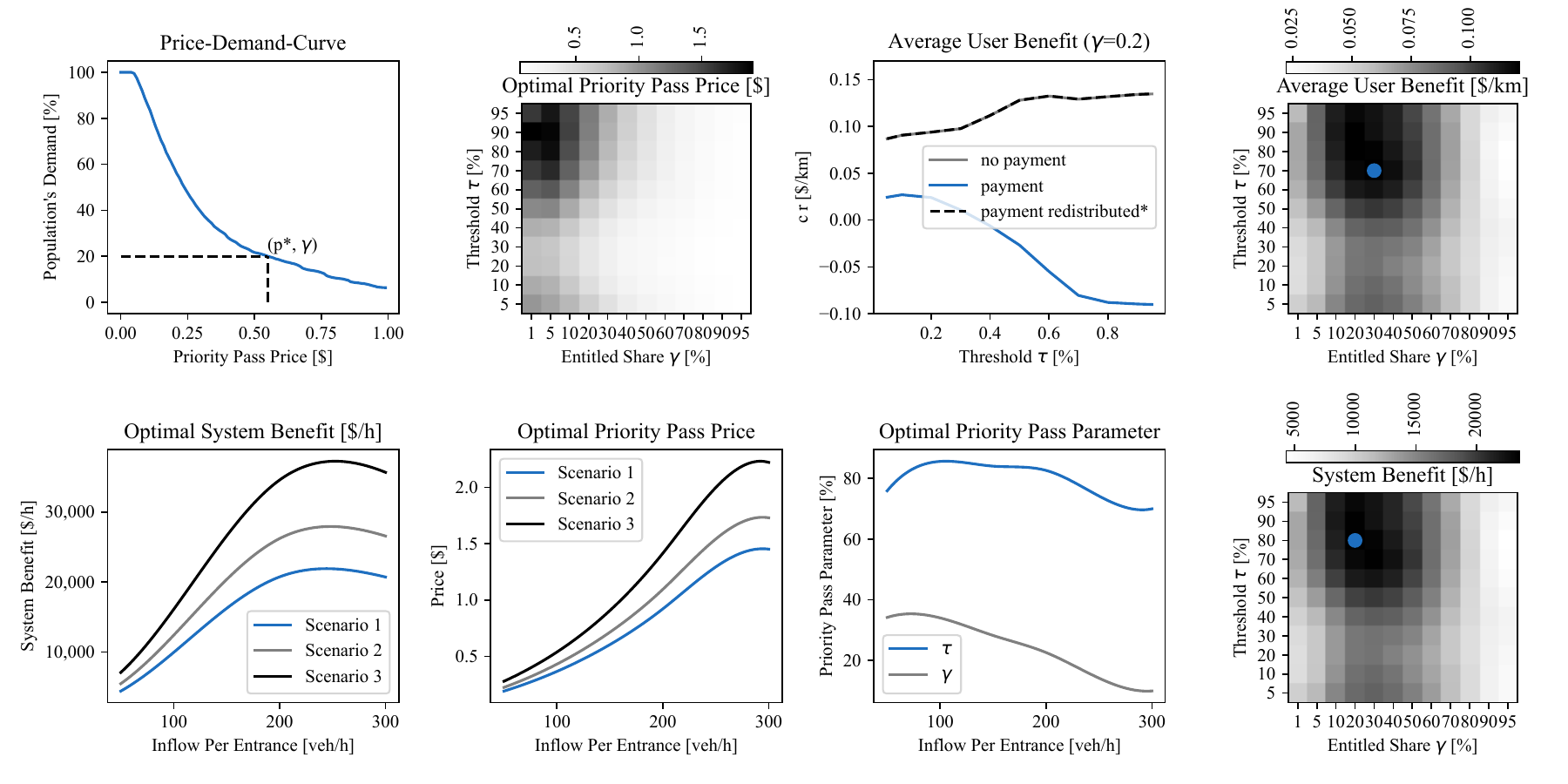}
   \caption{Optimization of the Priority Pass}
   \label{fig:fig3_optimization}
\end{figure*}

While simplifying assumptions on the VOT of users in the previous Fig.~\ref{fig:fig2_prioritypassmechanics} enabled an initial analysis, we now demonstrate the benefit of the Priority Pass given a realistic distribution of VOT based on the market model.
When optimizing the Priority Pass parameters $\gamma$ and $\tau$, two trade-offs need to be taken into account.
Fig~\ref{fig:fig3_optimization} showcases the optimization for the case study network.

The plots of the first row show different aspects of the optimization for an vehicle inflow per entrance to the case study network of 250 veh/h.
The first figure (from the left) shows how different prices for the Priority Pass result in different demand, and thus entitlement $\gamma$.
Consumers each have a specific income, urgency level, VOT, and a specific route to drive, with an expected travel time. 
Based on the expected delay changes when buying or not buying the Priority Pass, the consumers can calculate the opportunity costs of buying vs. not buying the Priority Pass.
These opportunity costs represent the reservation price of consumers.
With a growing price, less and less consumers have an opportunity cost greater than the market price, and therefore fewer consumers are willing to pay for the Priority Pass.
This relationship depends on the urgency level, income, route distribution of the population, and expected delay changes for a given $\tau$.
From a system-level perspective, the price is used as an instrument to control the entitlement rate $\gamma$ for a given $\tau$.
For instance, to achieve $\gamma=20\%$ a price of 0.55 \$ is suitable (e.g., for $\tau=50\%$).
The second figure (from the left) shows different optimal prices for the Priority Pass for different $\gamma$ and $\tau$ values.
The higher the price, the larger the reservation price and thus value of prioritization for the consumers.
The third figure (from the left) demonstrates the average user benefit $c_r$ for different $\tau$ assuming prices are chosen to guarantee $\gamma=20\%$.
The gray graph represents the case that the Priority Pass is granted to those individuals of the population with the largest VOT (largest need for prioritization) for free.
The allocation would require the system to know which individuals are those in highest need, and would rely on the honesty of all participants.
In this case around $0.1\$/km$ can be achieved, where generally larger $\tau$ lead to larger average user benefits $c_r$.
The few individuals that are prioritized save delay costs due to their expedition, at the cost of few more delay costs for all other individuals.
The average benefits are positive, and therefore the benefits of the prioritized individuals outweigh the disadvantages of the not entitled vehicles.
The blue graph represents the case that the Priority Pass is granted to those who are willing to buy it for the optimal price.
This approach would be more feasible, as individuals reveal their true, honest, VOT through their willingness to pay at most their reservation price.
While there are still regions where the Priority Pass can achieve average user benefits, in most of the cases it is not.
While the prioritized individuals enjoy fewer delay costs, they experience financial costs due to the price they need to pay. 
This is, why the benefits are significantly lower when compared with the free allocation discussed before.
The black dashed graph represents the case that the Priority Pass is granted to those who are willing to buy it for the optimal price, and all revenues are collected and equally distributed across all other individuals, that are not prioritized (entitled).
This results in similarly high average user benefits.
We advocate a Priority Pass based on a market resource allocation with a redistribution of revenues amongst the disadvantaged parts of the population.

Finally, the fourth figure (from the left) visualizes the average user benefit for different $\gamma$ and $\tau$ values. 
The colour of the heatmap (from white to black) represents the average user benefits that can be achieved (the darker the higher).
The blue dot represents the parameter combination $\gamma$ and $\tau$ that maximizes the average user benefit.
In this scenario, for $\gamma=30\%$ and $\tau=70\%$ the average user benefit reaches up to $0.125\$/km$.
The first trade-off is to prioritize entitled vehicles to such an extent, that it does not disadvantage the not urgent vehicles de trop.
The heatmap showcases that the average user benefit enables the optimization to satisfy this first trade-off. 
While the prioritization can enhance the average user's benefit and achieve higher levels of equity due to the better alignment of delays with the needs of individuals, it affects the transportation efficiency, by reducing flow and speed of the network. 
Thus, a second trade-off needs to be taken into account, to guarantee that the social, equitable welfare benefits for the average user do not affect the system's overall performance.
The heatmap on the right of the second row in Fig.~\ref{fig:fig3_optimization} shows the resulting system benefit for different $\gamma$ and $\tau$.
While the system benefit heatmap shares similarities with the average user benefit heatmap, the optimal parameter combination for this scenario lies at $\gamma=20\%$ and $\tau=80\%$.
While this solution is still generating significant average user benefits, this solution ensures the satisfaction of the second trade-off.
Average user benefit $c_r$ and system benefit $C_r$ are thus non-conflicting goals.
We advocate that the Priority Pass parameters are optimized for the system benefit, to justify few transportation efficiency losses to generate equity and welfare improvements.

The optimal Priority Pass parameters, and resulting system benefits and prices for the Priority Pass for different inflows per entrance into the network can be found in the second row of Fig.~\ref{fig:fig3_optimization}.
The two left figures showcase system benefit and price for the Priority Pass for different market model scenarios (urgency distributions).
The optimal parameters do not deviate too much for different market scenarios.
The optimal system benefits and generated social welfare reaches around 30,000 $\$/h$ for the small case study network consisting out of nine intersections.
The more heterogeneous (uneven) the urgency levels are distributed, the larger the possible welfare generations and the higher the value (price) of the Priority Pass to the individuals.
The Priority Pass parameters $\gamma$ and $\tau$ react sensitive to the level of saturation of the network, but remain around $\gamma=20\%$ and $\tau=80\%$.

\subsection{Priority Pass \& Effects on Transportation Efficiency}

\begin{figure*} [!htpb]
   \centering
   \includegraphics[width=0.98\linewidth]{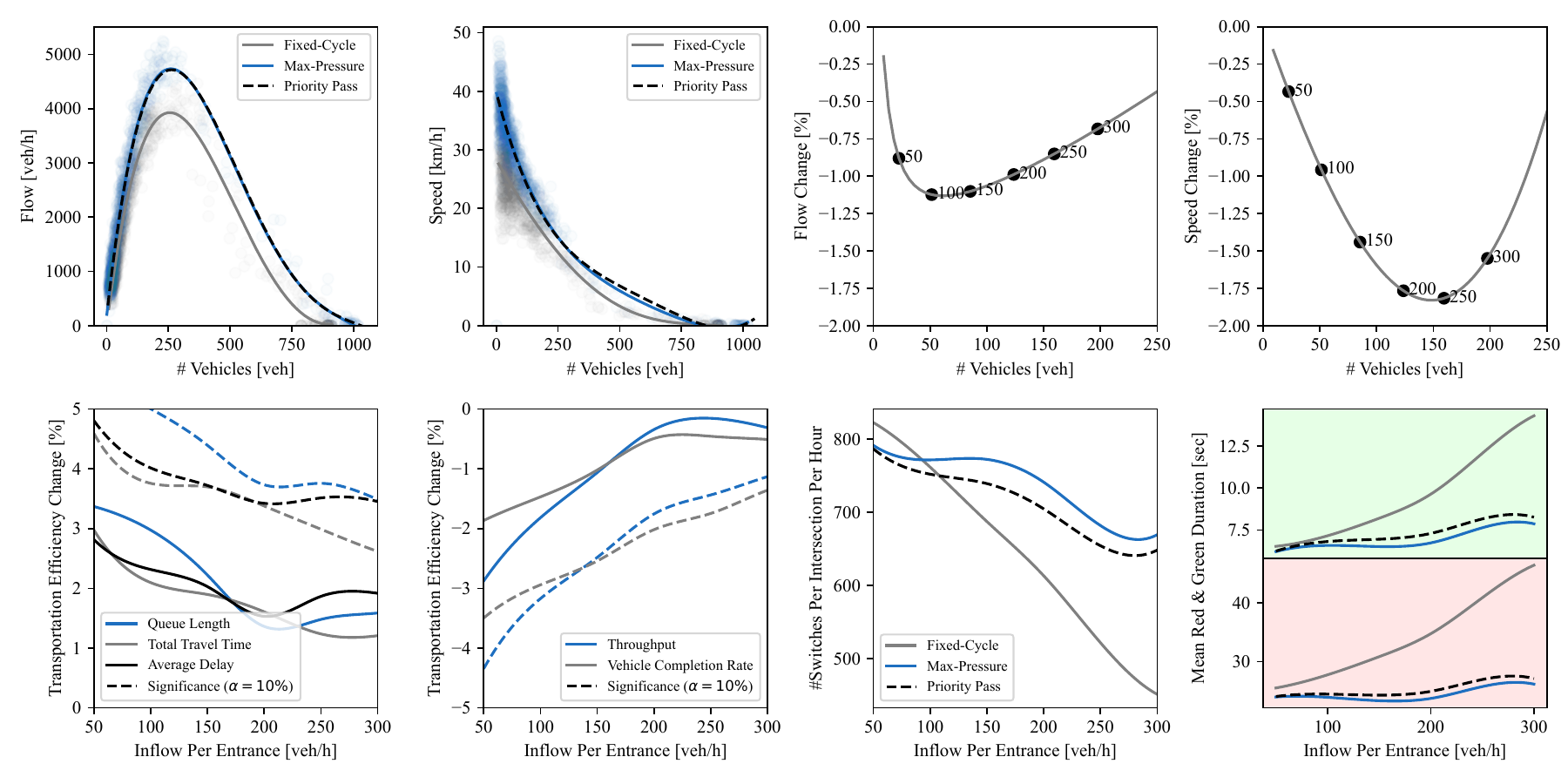}
   \caption{Transportation Efficiency Analysis of the Priority Pass}
   \label{fig:fig4_efficiency}
\end{figure*}

As discussed in the previous subsection, the Priority Pass was optimized to satisfy both trade-offs, considering the arbitrary effects on not-prioritized vehicles, and the potential losses of transportation efficiency.
Fig.~\ref{fig:fig4_efficiency} outlines the effects of the Priority Pass on transportation efficiency.

In the first row, conventional traffic fundamentals are explored to assess vehicle flow and average vehicle speed. 
The left two figures showcase the fundamental diagrams of the network for the three different traffic light control strategies fixed-cycle, Max-Pressure, and Priority Pass.
The fundamental diagram was recorded by ten simulations with different random seeds, where the fundamentals were recorded every 300 seconds. 
The simulations endured 30,000 seconds, and inflow per entrance was increased by 8.63\% every 1,000 seconds, starting from an inflow of 50 veh/h and reaching 600 veh/h during the course of the simulation.
The scattered dots represent fundamentals recordings, and the graphs represent a polynomial of fourth degree fitting the recordings.
The fundamental diagrams demonstrate, that the actuated, Max-Pressure controller is superior to the fixed-cycle controller, in that it achieves higher flows and speeds, and higher network capacities.
Moreover, the diagram exhibits, that the Priority Pass achieves similar traffic conditions.
The right two figures of the first row in Fig.~\ref{fig:fig4_efficiency} show the relative flow and speed change between Priority Pass and Max-Pressure controller.
The black dots on the curve represent the related inflows per lane that were used to achieve traffic states at this point of the curve.
The graphs demonstrate that the traffic fundamentals are both not reduced more than by 1.25\% (flow) and 1.75\% (speed). 
These reductions are not significant given the noise levels of the fundamentals recording.
Therefore, one can conclude that the prioritization of vehicles does not affect the transportation efficiency de trop.

Next, in the second row of Fig.~\ref{fig:fig4_efficiency}, more detailed traffic related evaluation measures are explored. 
The left figure shows how queue length, total travel time, and average delay are affected by the Priority Pass when compared with the Max-Pressure controller, for different inflows per entrance.
The dashed lines represent the significance intervals at a very generous significance level of 10\%.
Contrary to the fundamentals, these measures had even higher levels of variance and non-linearity.
The diagram demonstrates that the Priority Pass does not affect the transportation efficiency significantly.
Similarly, the second figure from the left demonstrates, that the Priority pass does not affect throughput and vehicle completion rate significantly.

Finally, an investigation on the traffic light signals was conducted in the second right plots, in the second row of Fig.~\ref{fig:fig4_efficiency}. 
The first plot shows the number of movement phase switches per intersection and hour.
The fixed-cycle controller switches phases fewer times when compared with the actuated controllers.
The Priority Pass switches phases slightly fewer when compared with the Max-Pressure controller.
The larger the inflow per entrance, the lower the number of times switches are happening.
The second plot shows the average red \& green durations per signal.
The average green and red phase durations of the fixed-cycle controller are higher than those of the actuated controllers, which is plausible, as the number of times movement phase switches happen is lower.
The Priority Pass obtains average green and red phase durations similar to those of the Max-Pressure controller.
The larger the inflow per entrance, the longer the durations become, which is plausible, as the number of times movement phase switches happen decreases.

\begin{figure*} [!htbp]
   \centering
   \includegraphics[width=0.98\linewidth]{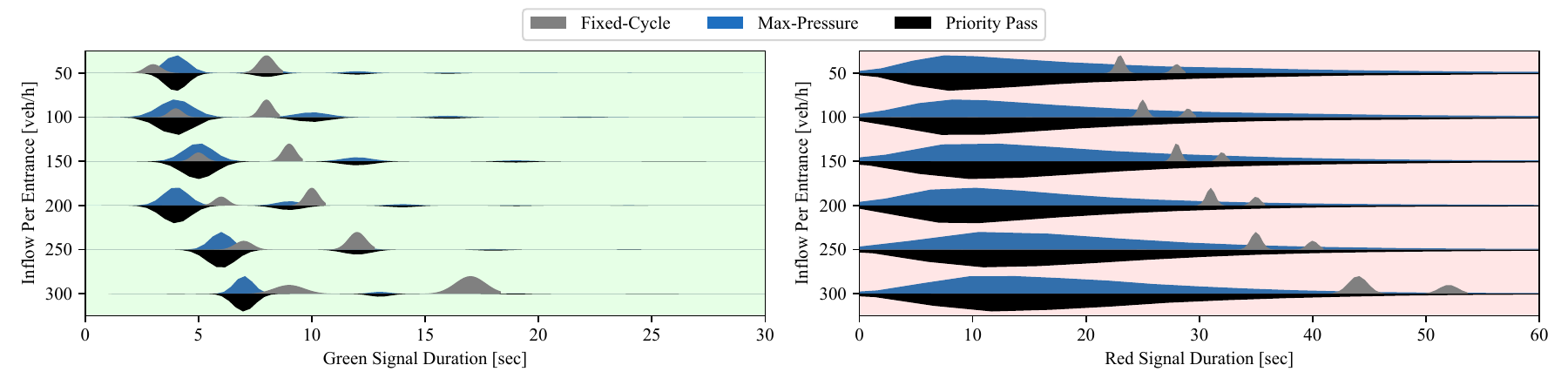}
   \caption{Traffic Light Signal Analysis of the Priority Pass}
   \label{fig:fig5_tls}
\end{figure*}

A more in-depth analysis of the distribution of green and red phase durations can be found in Fig.~\ref{fig:fig5_tls}.
The distributions are displayed as violin plots, representing the probability distribution of green and red signal durations.
The diagrams showcase, that the actuated controllers have a stronger variation of green and red signal durations.
Especially the red signal durations posses a stronger variation when compared to the green signal durations.
Moreover, the fixed-cycle controller's distributions obtain two significant peaks, representing the two green phase durations that were used for horizontal and vertical movements.
The variance in the diagrams exemplify, that actuated controllers react to traffic conditions, while fixed-cycle controllers do not.

\subsection{Fair \& Efficient Allocation of the Priority Pass}

\begin{figure*} [!htpb]
   \centering
   \includegraphics[width=0.64\linewidth]{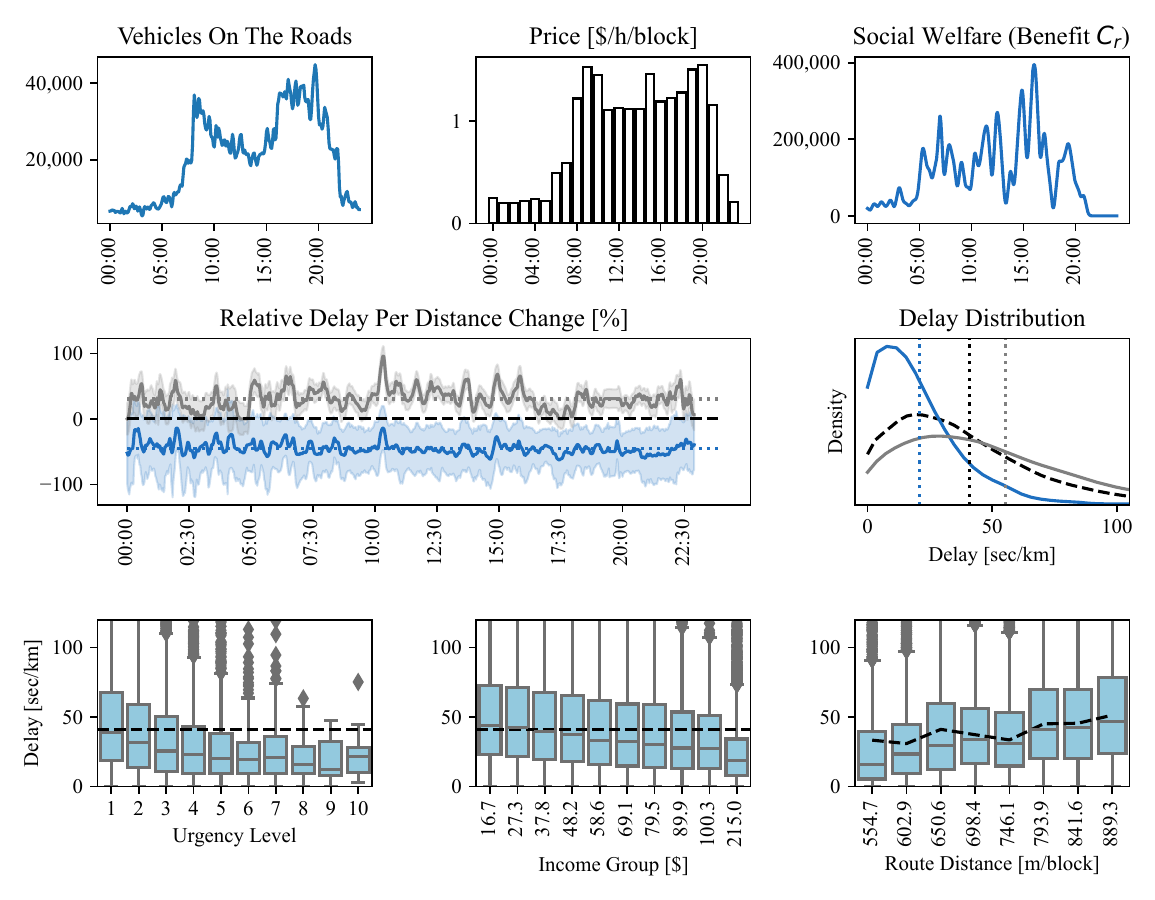}
   \includegraphics[width=0.34\linewidth]{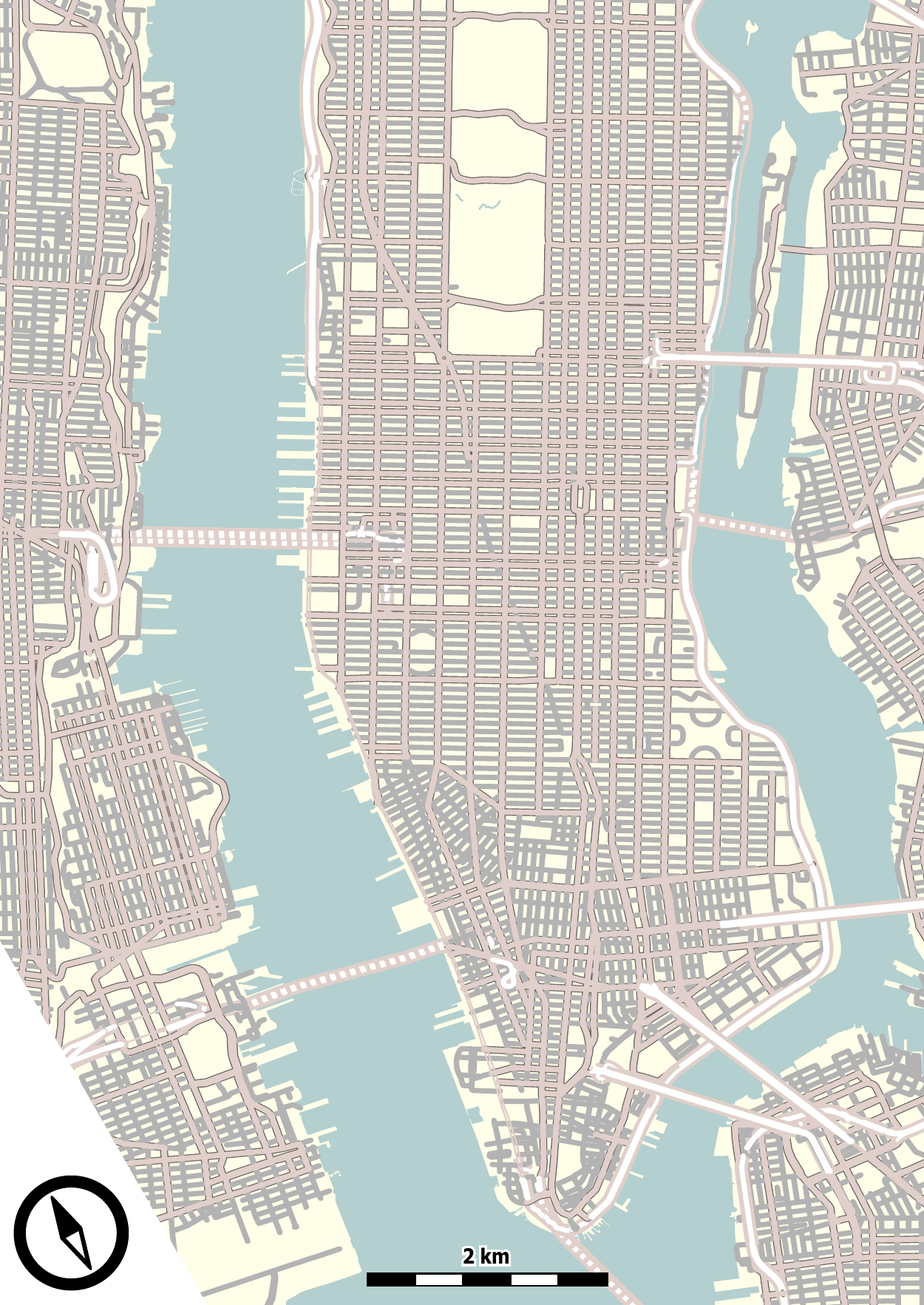}
   \caption{Social Welfare Analysis of the Priority Pass For Manhattan}
   \label{fig:fig6_welfare}
\end{figure*}

In this part of the analysis, we aim to quantify the potential social welfare benefit that could be generated by the Priority Pass for a network as big as Manhattan, with more than 2,862 intersections and a daily travel volume of 5,958,060 vehicle trips (on average 496,505 veh trips/h), for a whole day (24 h).
Fig.~\ref{fig:fig6_welfare} presents the results of this exercise.
On the right, a map illustrates the borough of Manhattan in New York City (USA).

\begin{table}[!hbp]
    \centering
    \begin{tabular}{|lrrr|}
        \hline
        \textbf{ } & \multicolumn{3}{c|}{\textbf{Urgency Scenarios}} \\
        \textbf{ } & \textbf{1} & \textbf{2} & \textbf{3} \\
        \hline
        \textbf{Welfare Generated $C_r$} & 95.53 & 126.42 & 158.80 \\
        \;\;\; {[Mio. \$ per day]} &   &   &   \\
        \textbf{Welfare Generated $c_r$} & 16.03 & 21.22 & 26.65 \\
        \;\;\; {[\$ per user \& user]} &   &   &   \\
        \textbf{Prices} & 0.84 & 1.00 & 1.29 \\
        \;\;\; {[\$ per h \& block]} &   &   &   \\
        \textbf{Revenues} & 875.96 & 1,041.16 & 1,343.78 \\
        \;\;\; {[Thsd. \$ per day]} &   &   &   \\
        \hline
        \textbf{Total Flow } & \multicolumn{3}{l|}{5,958,060} \\
        \;\;\; {[veh per day]} & \multicolumn{3}{c|}{ } \\
        \textbf{$\gamma$ Prioritized Population } & \multicolumn{3}{l|}{23.3} \\
        \;\;\; {[\%]} & \multicolumn{3}{c|}{ } \\
        \textbf{\# Prioritized Population } & \multicolumn{3}{l|}{42,586} \\
        \;\;\; {[user per day]} & \multicolumn{3}{c|}{ } \\
        \hline
    \end{tabular}
    \caption{Welfare Analysis \& Comparison of Scenarios}
    \label{tab:scenario_comparison}
\end{table}


The daily traffic pattern that was used for simulation of New York (according to \citealp{chaloli2019paradigmatic}) is shown in the plot on the top left.
The graph shows the number of vehicles on the road at the same time for every 300 seconds.
The second plot in the first row shows the price for the Priority Pass over the day.
A reservation per hour is assumed, which is why the parameters (price, $\gamma$, $\tau$) were optimized on an hourly level.
During peak hours, the Priority Pass price can go as high as 1.5\$ for an hour per block (nine intersections).
The third plot in the first row shows the social welfare $C_r$ in \$/h for every 300 seconds.
At peak times the prioritization can save up to 400,000\$/h for the whole driver population when compared to a situation without prioritization.
The second and third plot were calculated for urgency scenario 1, assuming a redistribution of generated revenues.
For the other two scenarios, similar curves over the day could be observed.

The second row of Fig.~\ref{fig:fig6_welfare} showcases the delay distribution for prioritized (blue curve, entitled with Priority Pass) and not-prioritized vehicles (gray curves), when compared to the situation without prioritization (black dashed curve).
On average (for the whole day), a prioritized vehicle saved up to 45.04\% delays (per distance), while not-prioritized vehicles gained up to 30.55\% delays (per distance).
The right figure of the second row shows the delay (per distance) distribution for the three different groups (aggregated for the whole day).
The plot showcases, that the prioritized vehicles (blue dotted curve) have a significantly lower delay on average, when compared to the situation without a Priority Pass (black dotted curve), while the not-prioritized vehicles have a slightly higher delay on average (gray dotted curve).
Moreover we found that the standard deviation of delays for the prioritized vehicles are greater when compared with the not-prioritized vehicles.

The third row of Fig.~\ref{fig:fig6_welfare} analyses the delay distribution for different urgency levels, income groups and route distances as boxplots, and compares those with the situation without prioritization (black dashed curve).
The first plot shows, how the Priority Pass enhances the fairness of the transportation system, as mobility resources are allocated to those users most in need, which causes significant delay reductions for higher urgency levels.
Especially for all users with an urgency level greater than 4, the inter-quartile-range sinks below the average delay in a situation without prioritization.
In addition to that, the second plot shows, that the Priority Pass is aligned with the needs of users, rather than with the income (wealth) of individuals. 
For sure, wealthier individuals achieve lower delays on average, but with regards to the whole driver population, only those with a significantly higher income (greater than 215\$/h) achieve the inter-quartile-range to sink below the average delay in a situation without prioritization.
The third plot shows, that the Priority Pass does not affect the delay distribution over route distances. The boxplots (Priority Pass) show a similar pattern to the black dashed lined (control without prioritization).
To summarize, even a decentralized, market-like allocation scheme could achieve an equitable distribution of the mobility resources to those most in need.

Table~\ref{tab:scenario_comparison} quantifies the welfare analysis for the three different urgency distribution scenarios.
Over the course of the whole day, prioritization with the Priority Pass achieved over 95,529, 755 \$ (scenario 1), 126,421,543 \$ (scenario 2), respectively 158,799,328 \$ (scenario 3) in social welfare benefits for the whole day and the whole network of Manhattan, this accounts for almost 16.03\$ (scenario 1), 21.22\$ (scenario 2), respectively 26.65\$ (scenario 3) welfare benefits per average user for the day.
Moreover, revenues of 875,960\$ (scenario 1), 1,041,160\$ (scenario 2), respectively 1,343,780\$ (scenario 3) could be generated for the municipal authorities.
The generated welfare by prioritization is higher for more unequal distributions of urgency, which is plausible as the core idea of the Priority Pass is to generate welfare benefits by not treating user equally as not every user experiences the same costs when being delayed.

This welfare benefit could be interpreted in several ways.
First, assuming the Priority Pass is allocated based on honest, self-reported urgency, these benefits could achieve more equitable, needs-oriented allocation of mobility resources.
Second, assuming Priority Pass holders to pay and a redistribution of the fees for prioritization, these benefits could similarly help achieve social goals, as well as environmental and economic goals.
For instance, this welfare benefits could increase the profits of commercial vehicles such as taxis and delivery vehicles, which gains higher tax revenues for the city.
Moreover, as commercial vehicles are expedited, less commercial vehicles might be necessary to cover the same number of trips, which could induce a traffic demand reduction on the long-term.
Considering that around 15\% of vehicles on Manhattan's road are delivery vehicles, and around 7\% of vehicles account for taxis, the potential for traffic demand reduction is significant.
This potential traffic demand reduction could help achieve less traffic and thus environmental goals.
Finally, assuming the road authorities collect the revenues instead of redistributing them, additional revenues for the city to invest into the road infrastructure and public transportation of 875,960\$ (scenario 1), 1,041,162\$ (scenario 2), respectively 1,343,780\$ (scenario 3) per day could be generated. 
Manhattan already generates daily revenues of around 6,570,000\$ through \href{https://new.mta.info/press-release/mta-bridges-and-tunnels-announces-record-revenue-2023-enforcement-metrics-increase}{tunnel and bridge tolls}, and will soon generate even further revenues through \href{https://edition.cnn.com/2025/01/12/business/nyc-congestion-pricing-2025/index.html}{congestion pricing}. Therefore, the revenue potential for a Priority Pass could help to significantly increase revenues for traffic authorities in Manhattan and New York City.




\section{Conclusions}
\label{sec:conclusion} 

Even though they are the major bottleneck for traffic in cities, no dedicated instrument addresses the need for prioritization at intersections.
In this work we set out to propose a novel, feasible, economic instrument to address needs-oriented signalized intersection management in cities: the Urban Priority Pass.
A case study of Manhattan demonstrates that the prioritization of few vehicles is possible without significantly affecting transportation efficiency or causing arbitrary delay to others.

The results of the case study advocate the welfare benefits of the Priority Pass; there are five key takeaways from this study: 
First, it is possible to expedite entitled vehicles by prioritizing them in traffic light control.
Second, it is possible to expedite entitled vehicles without causing arbitrary, disproportionate delays for not-entitled vehicles.
Third, the benefits of the Priority Pass do not come at the cost of transportation efficiency.
Fourth, significant benefits for system and users are observed, with delay reductions up to 40\% for entitled vehicles. For a network of the size of Manhattan, prioritization could create up to 100 million \$ per day in welfare, and more than 1 million \$ revenue per day for the city that could be used to improve public transportation and maintain road infrastructure.
Fifth, a market for the Priority Pass systematically benefits users of higher levels of urgency, rather than individuals of higher income. 

By aligning delays with the needs of drivers through prioritization at intersections, the Priority Pass has the potential to achieve social, environmental and economic goals, and enhances the fair allocation of scare, spatio-temporal mobility resources.

In future works, one could explore the user acceptance of a Priority Pass by private and commercial drivers through surveys.
One could assess the willingness to pay for prioritization and urgency level distributions to estimate prices, market sizes, and potential revenues for municipalities.
The estimation of the economic value for taxis or delivery vehicles based on available trip datasets could be a promising research direction.
Moreover, one could explore the potential for traffic demand reduction through prioritization, as commercial vehicles could achieve the same number of trips with fewer vehicles due to significantly shorter delays.
Additionally, one could explore how prioritization of vulnerable road users might enhance their road safety.
Also, one could explore disruption- and turbulence-free prioritization for transit signal priority and emergency vehicle preemption using the Priority Pass.
Besides, one could explore not one, but many different levels of prioritization.
Finally, the Priority Pass could be applied in the context of multimodal transportation, where it could enable the mobility resource allocation for prioritization of dedicated transportation modes.




\renewcommand{\bibfont}{\scriptsize}
\bibliographystyle{elsarticle-harv} 
\bibliography{main_bibliography.bib}

\end{document}